\documentstyle[11pt]{article}

\begin{document}

\hoffset=-20mm
\voffset=-8pt

\vspace*{26mm}

{\Large\bf
\noindent
On the Available Lunar and Solar Eclipses and Babylonian Chronology}\\[6mm]
{\large V.G.Gurzadyan$^1$ and D.A. Warburton$^2$}\\[6mm]
{(1) ICRA,  University  of  Rome ``La Sapienza", Italy  and   Yerevan  Physics
Institute, Armenia; (2)	University of Aarhus, Denmark}\\[5mm]

(Published in {\it Akkadica},  v.126 (2005) p.195.)	 

\vspace{0.1in}

The recently shown two premises (Gurzadyan 2000), i.e. the absence of 56/64 year Venus cycle constraints, at the importance of the 8-year cycle in the Venus Tablet, stimulated new studies on the Chronology of the Ancient Near East (2nd millennium BC). The analysis by B.Banjevic using both premises, however, did not provide anchors of strenght similar to those of Ur III eclipses, while available solar eclipses lack unambiguous links to historical events. The Ultra-Low chronology (Gasche et al 1998), therefore, has to be considered as currently the one most reliably based on ancient astronomical records.  

\vspace{0.2in}

There are two premises at the base of Boris Banjevic's (2005) chronological analysis.  The first is that chronological proposals can no longer be legitimately constrained by 56/64 Venus-cycles (which are the base of the High, Middle and Low chronologies previously recognized), as these cycles cannot be reliably established in the existing sources.  The second premise is that the 8-year cycles can be extracted from the Venus Tablet. Both premises were shown in our work (Gurzadyan 2000).

Based on the historical records, the Ultra-Low chronology (Gasche et al 1998) satisfied the 8-year Venus cycle condition, which is recognized by all concerned.  Furthermore, this was independently anchored via the lunar eclipses, and thus with complementary astronomical sources linked to the condition of the interval between the eclipses known from historical sources.  It was the final phase information of the records of the Ur III lunar eclipses of EAE 20 IIIA and EAE 21 XII, including both the watch times and the additional condition in EAE 20 IIIB about the "weakly shining stars" (showing the absolute importance of the exit information), which allowed the chronology to be fixed precisely.

In contrast to this procedure, Banjevic does not provide any new anchors of similar strength.  To discuss $\Delta$T in order to fit the lunar eclipse of -1960 of twilight to the second Ur III eclipse seems unreasonable in principle, particularly given the absence of an alternative link to historical events.  Furthermore, e.g., the recently reported lunar eclipse of -382 seems to pose new problems for our understanding of $\Delta$T (Steele, 2005).  Thus this subtle issue must be approached  cautiously.

Individual solar eclipses, as discussed previously in the literature, can hardly act as anchors, without unambiguous information associating to history.  Similarly, the Assur Solar eclipse of -1764 even though compatible to Ultra-Low chronology (unaffected by $\Delta$T!), has never been used as a crucial argument for the latter.

As an astrophysicist, one of us (V.G.) only commented on astronomical aspects, where it would seem that additional astronomical support compatible with the historical givens should be expected.

Making a contribution in this respect requires an interdisciplinary effort combining a profound knowledge of texts, language, religion, history and archaeology.  Thus, caution might be advised when placing too much confidence in statistical probabilities aligned with some possible correlations.  Banjevic highlights the difficulties in noting that even according to his own interpretation (which need not correspond to the calculations of others), his proposal involves a discrepancy of "only 3 years" with respect to the Assyrian Distanzangaben. In fact, however, the size of the gap is immaterial, as it demonstrates (as Eder 2004 has already shown) that the Distanzangaben are not compatible with any known astronomical data - even Banjevic's.

\renewcommand{\baselinestretch}{1}
\section*{\it References}

\begin{description}

\item[ ]\hspace{-1mm} BANJEVIC, B., 2005, {\it Akkadica}, 126, 169.  
    \\[-7mm]
\item[ ]\hspace{-1mm} EDER, Ch., 2004 : "Assyrische Distanzangaben und die Chronologie Vorderasiens", {\it Altorientalische Forschungen} 31, 191-236.  
    \\[-7mm]
\item[ ]\hspace{-1mm} Gasche et al 1998 = GASCHE, H., ARMSTRONG, J.A., COLE, S.W., GURZADYAN, V.G., 1998 : {\it Dating the Fall of Babylon.} A Reappraisal of Second-Millennium Chronology (= MHEM 4), Ghent, Chicago.  \\[-7mm]

\item[ ]\hspace{-1mm} GURZADYAN, V.G., 2000 : On the Astronomical Records and Babylonian Chronology, Just in Time. Proceedings of the International Colloquium on Ancient Near Eastern Chronology (2nd Millennium BC). Ghent 7-9 July 2000 (= {\it Akkadica} 119-120), Bruxelles, 177-186; physics/0311035.  
    \\[-7mm]
\item[ ]\hspace{-1mm}  STEELE, J., 2005 : Ptolemy, Babylon and the Rotation of the Earth, {\it Astronomy \& Geophysics} 46, 5.11-5.15.\\[-7mm]

\end{description}

\end{document}